\documentclass[11pt]{article}
\usepackage{amsmath, amssymb, amsthm}
\newtheorem{theorem}{Theorem}
\newtheorem{lemma}{Lemma}

\newtheorem{corollary}{Corollary}

\theoremstyle{definition}
\newtheorem{remark}{Remark}

\def \whrho {\widehat{\rho}\ }

\def \wt {\widetilde}

\def \FF  {{\cal F}}
\def \GG  {{\cal G}}
\def \HH  {{\cal H}}

\def \VV  {{\cal V}}
\def \WW  {\mathcal W}

\def \CCC {\mathbb{C}}

\def \KKK {\mathbb{K}}
\def \NNN {\mathbb{N}}
\def \PPP {\mathbb{P}}

\def \RRR {\mathbb{R}}
\def \ZZZ {\mathbb{Z}}

\def \FFI {\widetilde{\cal F}}

\newcommand \tr {\operatorname{Tr}}
\newcommand \Tr {\tr}

\newcommand \spann {\operatorname{span}}
\newcommand \Ub {\mathbf{U}}
\newcommand\hU{\widehat{U}}
\newcommand \Lx {\operatorname{L}}
\newcommand\slim{\operatornamewithlimits{s\cdot\lim}}
\newcommand\dotsv{V_1,\dots,V_d}
\newcommand{\Domega}{D(\omega)}
\newcommand{\pkj}{P_{kj}}
\newcommand{\ddo}{\partial_{\omega}}

\title{Convergence of coined quantum walks on $\RRR^d$}
\begin{document}

\author{
Alex D. Gottlieb%
\footnote{ Wolfgang Pauli Institute c/o Institut f\"ur Mathematik,
Universit\"at Wien, Nordbergstra\ss e 15, 1090 Wien, Austria
(alex@alexgottlieb.com). },
Svante Janson%
\footnote{ Department of Mathematics, Uppsala University, PO Box
480, S-751 06 Uppsala, Sweden (svante.janson@math.uu.se).}, and
Petra F. Scudo%
\footnote{ Department of Physics, Technion---Israel Institute of
Technology, 32000  Haifa, Israel (scudo@tx.technion.ac.il). }
}

\date{June 7, 2004}

\maketitle

\abstract{ Coined quantum walks may be interpreted as the motion
in position space of a quantum particle with a spin degree of
freedom; the dynamics are determined by iterating a unitary
transformation which is the product of a spin transformation and a
translation conditional on the spin state. Coined quantum walks on
$\ZZZ^d$ can be treated as special cases of
  coined quantum walks on $\RRR^d$.
We study quantum walks on $\RRR^d$ and prove that the sequence of
rescaled probability distributions in position space associated to
the unitary evolution of the particle converges to a limit
distribution.  }

\section{Introduction}

Several kinds of quantum walks have recently been studied by many
authors; a nice overview is given by \cite{Kempe}.  So-called
``coined quantum walks'' on finite graphs, introduced in
\cite{AAKV}, are proving to be of some interest in quantum
informatics, where they have been used to devise fast search
algorithms \cite{Ambainis,Kempe,SKW}.   Quantum walks of the type
considered in this article, namely, coined quantum walks on
$\ZZZ^d$ or $\RRR^d$, were first introduced in \cite{ABNVW},
though \cite{ADZ} and \cite{Meyer} can be considered as precursors
in some respects.

A coined quantum walk describes the evolution of a quantum system
under iteration of a certain kind of unitary map. The state of the
system is a vector in a product Hilbert space, having position
degrees of freedom and an internal degree of freedom such as spin
or polarization. A step of the walk consists of a unitary
transformation of the spin degree of freedom, which one may think
of as ``tossing a quantum coin'', followed by a translation
conditional upon the state of the coin. Here we restrict our
attention to walks on $\ZZZ^d$ or $\RRR^d$ and we prove that the
position distribution of the quantum walker, properly rescaled,
converges as the number of steps tends to infinity.  This type of
convergence was first discovered by N. Konno \cite{K1,K2} in the
case of one-dimensional lattices $\ZZZ$. A different proof of
Konno's theorem was given in \cite{GJS}, generalizing the
convergence to quantum walks on higher dimensional lattices
$\ZZZ^d$. In this article we present a further generalization of
the result to quantum walks on $\RRR^d$, and we obviate a
technical difficulty of \cite{GJS} that required an extra
hypothesis there.

\section{Basic definitions and review of Konno's theorem}
\label{S:basic}

We start by reviewing some basic definitions and stating Konno's
theorem for a coined quantum walk on $\ZZZ$.

A simple
quantum walk on $\ZZZ$ can be defined as a sequence
\begin{equation}
\label{orbit}
     \psi,\ U\psi,\ U^2\psi,\ U^3\psi,\ \ldots,
\end{equation}
of unit vectors in the Hilbert space $\ell^2(\ZZZ)\otimes \CCC^2$
obtained by iterating a unitary operator $U$ of the form
\begin{equation}
\label{special}
     U=S(I\otimes C)\ .
\end{equation}
In (\ref{special}), $I$ denotes the identity operator on
$\ell^2(\ZZZ)$, $C$ denotes a unitary operator on $\CCC^2$, and
$S$ denotes the ``conditional shift'' operator
\begin{eqnarray}
        S( s_m \otimes e_1 )  & = & s_{m+1}  \otimes e_1 \nonumber \\
        S( s_m \otimes e_2 )  & = &  s_{m-1} \otimes e_2
        \label{conditionalShift} \ ,
\end{eqnarray}
where $e_1$ and $e_2$ are the standard basis vectors for $\CCC^2$
and $s_m$ is the vector in $\ell^2(\ZZZ)$ whose $m^{th}$ member is
$1$ and all other members are $0$. Since the vectors of the
sequence (\ref{orbit}) are normalized, the numbers
\begin{equation}
\label{probabs}
      P_n(m) \ = \ \big|\langle s_m \otimes e_1,\ U^n\psi \rangle\big|^2
      + \big|\big\langle s_m \otimes e_2,\ U^n\psi \rangle\big|^2
\end{equation}
satisfy $\sum_m P_n(m)=1$ and thus define
a sequence of
probability measures on $\ZZZ$. Note that the only variable
considered here is the position of the walker; the internal degree
of freedom is disregarded by summing over both states in the coin
space. Konno's theorem \cite{K2} states that the probability
measures
\begin{equation}
\label{Konno}
         \sum_{m \in \ZZZ} P_n(m)\ \delta_{m/n}
\end{equation}
converge weakly as $n \longrightarrow \infty$ to a probability
measure that depends on the initial state $\psi$ (here
$\delta_{m/n}$ denotes a point-mass at $m/n$).

The dynamics described in (\ref{orbit}), (\ref{special}),
(\ref{conditionalShift}) is reminiscent of simple random walk on
$\ZZZ$, though there are a few important differences. As in
ordinary random walks, the rule (\ref{conditionalShift}) for
stepping left or right is the same at all locations $m \in \ZZZ$,
that is, the process is spatially homogeneous. Unlike ordinary
random walks, the probability distributions (\ref{probabs}) for
the system's position are not related by Markov transitions.
Similarly to random walk, the sequence of rescaled position
distributions (\ref{Konno}) converges weakly, but unlike random
walks, the weak limit is obtained by rescaling by a factor of
$1/n$ rather than $1/\sqrt{n}$ and the limit distribution depends
on the initial state (and is not
a normal distribution).

In the next sections we state and prove a generalization of
Konno's theorem in which (i) the walk takes place in
$d$-dimensional space $\RRR^d$, (ii) any finite number $s$ of
conditional translations are allowed at each step, and (iii) the
translations are not assumed to generate a lattice in $\RRR^d$.
Quantum walks on the $d$-dimensional lattice $\ZZZ^d$ may be
viewed as special cases of quantum walks on $\RRR^d$ where the
shifts generate a discrete lattice.

\section{General quantum walks convergence theorem}
\label{S:gen}

Broadly speaking, a quantum walk is characterized by the unitary
operator $U$ that generates the walk, $U$ being a particular kind
of unitary operator on a Hilbert space $L^2(\ZZZ^d)\otimes \CCC^s$
or $L^2(\RRR^d)\otimes \CCC^s$, the former for quantum walks on
$\ZZZ^d$ and the latter for quantum walks on $\RRR^d$.  We now
focus our discussion onto quantum walks on $\RRR^d$, for quantum
walks on $\ZZZ^d$ are easily embedded into the framework of walks
on $\RRR^d$.

Let $\HH = L^2(\RRR^d)\otimes \CCC^s$.  There is a natural
isomorphism between $L^2(\RRR^d)\otimes \CCC^s$ and $L^2(\RRR^d,
\CCC^s)$; it will be helpful to represent members of $\HH$ as
members of $L^2(\RRR^d, \CCC^s)$, and indeed to represent members
of $\CCC^s$ as column vectors for the purpose of interpreting
matrix operations. Let $C$ denote a unitary ``coin tossing
operator'' on the ``coin space'' $\CCC^s$ and let $I_s$ denote the
identity operator on $\CCC^s$. Let $e_1,e_2,\ldots,e_s $ be the
standard ordered basis of $\CCC^s.$ The steps of the quantum walk
involve a ``conditional translation'' operator $T$ on $\HH$
defined by
\begin{equation}
\label{conditionalTranslation}
         T( \psi \otimes e_j ) \ = \ T_{v_j}\psi \otimes e_j\ ,
\end{equation}
where $T_{v_j}$ denotes translation by $v_j$ in $\ZZZ^d$ or
$\RRR^d$, i.e.,\ $(T_{v_j}\psi)(x) = \psi(x - v_j) $.
The vectors $v_j$ are arbitrary fixed vectors in $\ZZZ^d$ or
$\RRR^d$. A step of the quantum walk is effected by a unitary
operator
\begin{equation}
\label{step}
    U \ = \ T (I\otimes C),
\end{equation}
where $I$ denotes the identity operator on $L^2(\ZZZ^d)$ or
$L^2(\RRR^d)$.

The operator $U$ generates the quantum walk.   In the
Schr\"odinger picture, the quantum walk may be thought of as a
sequence
\begin{equation}
\label{orbit2}
     \rho,\ U\rho U^*,\ U^2 \rho U^{*2},\ U^3 \rho U^{*3},\ \ldots
\end{equation}
of density operators on $\HH$.  (Recall that a density operator is
a nonnegative trace class operator with normalization
$\Tr(\rho)=1$ \cite{CCT}. The density operator $\rho$ for the
normalized pure state $\psi \in \cal{H}$ has integral kernel
$\psi(x) \overline{\psi(y)}$. Thus, the definition (\ref{orbit2})
of quantum walk in terms of density operators generalizes the
original definition (\ref{orbit}) for pure states $\psi$.)  The
probability measures (\ref{probabs}) can be expressed in terms of
density operators instead of wavefunctions as follows. Given an
initial density operator $\rho$ on $\HH$ and any $n \in \NNN$,
there is a probability measure $\PPP_{\rho,n}$ on $\ZZZ^d$ or
$\RRR^d$ that may be defined as a linear functional on
$C_0(\ZZZ^d)$ or $C_0(\RRR^d)$ by the formula
\begin{equation}
\label{unscaled}
        \PPP_{\rho,n}f \ = \  \tr( U^n \rho U^{*n} ( m[f] \otimes
        I_s)),
\end{equation}
where $C_0$ denotes the set of continuous functions that tend to
$0$ at $\infty$, $f$ is an element of this set, and $m[f]$ is the
operator of multiplication by $f$. Formula (\ref{unscaled}) may be
written as
\begin{equation}
\label{partialTraced}
        \PPP_{\rho,n}f \ = \  \tr( \tr_{\CCC^s}(U^n \rho U^{*n}) m[f] ),
\end{equation}
where $\tr_{\CCC^s}$ denotes the partial trace; for the special
case studied in Section~\ref{S:basic}, one may verify that
$\PPP_{\rho,n}$ is the same as $P_n$ of (\ref{probabs}) when
$\rho$ is a pure state $\psi$. The quantum walk as such is the
sequence of density operators (\ref{orbit2}) and the corresponding
probabilities (\ref{partialTraced}), which express probabilites
concerning the position of a ``particle'' without regard to its
internal ``spin'' state in $\CCC^s$.

One observes that the probability measures $\PPP_{\rho,n}$ spread
out in $\ZZZ^d$ or $\RRR^d$ as $n$ increases.  However, we will
discover that dilating each of the $\PPP_{\rho,n}$ by a factor of
$1/n$ produces in a sequence of probability measures that
converges weakly.  In other words, if $X_n$ is a random vector
describing the position of the quantum
walk after $n$ steps, 
then the sequence $X_n/n$ converges in distribution.

Now we are prepared to state a generalized version of Konno's
theorem \cite{K1,K2,GJS}.   In the following, $C_b(\RRR^d)$
denotes the space of bounded continuous functions on $\RRR^d$, and
a sequence of probability measures is said to converge weakly if
the measures converge pointwise as functionals on $C_b(\RRR^d)$
\cite{Billingsley}.

\begin{theorem}
 \label{theTheorem} Let $U$ be a unitary operator on $\HH =
L^2(\RRR^d)\otimes \CCC^s$ of the form \eqref{step}.

There exists a spectral measure $P(dx)$ on $\RRR^d$ such that for
any density operator $\rho$ on $\HH$, the probability measures
$\PPP^{(n)}_{\rho,n}$ defined by
\begin{equation}
\label{together}
       \PPP^{(n)}_{\rho,n}f(x) \ = \  \Tr\bigl(
       \Tr_{\CCC^s} (U^n \rho U^{*n})
       m[f(x/n)] \bigr)
\end{equation}
converge weakly as $n \longrightarrow \infty$ to the probability
measure $\PPP_{\rho}(E)=\tr(\rho P(E))$.
\end{theorem}

We will prove this theorem --- and somewhat more --- in the next
section, but first it is appropriate to explain in which sense
Konno's theorem for walks on the lattice is generalized by
Theorem~\ref{theTheorem}.  Convergence of quantum walks on
$\RRR^d$ implies convergence of quantum walks on $\ZZZ^d$ because
quantum walks on $\ZZZ^d$ can be identified with quantum walks on
$\RRR^d$ that have integral steps.  Indeed, the space
$L^2(\ZZZ^d)$ can be identified with the subspace $\KKK \subset
L^2(\RRR^d)$ consisting of functions that are constant on each
unit cube centered at a lattice point. The subspace $\KKK$ is
invariant under translation by any lattice vector in $\ZZZ^d$, so
that a quantum walk on $\ZZZ^d$ may be viewed as a quantum walk on
$\RRR^d$ whose initial density operator is supported on the
subspace $\KKK \otimes \CCC^s$ of $L^2(\RRR^d) \otimes \CCC^s$.
The analog of an initial state localized at a vertex of the
lattice is a state supported on a cube centered at the
corresponding lattice point.  The limit distribution for the
discrete case can be retrieved by taking the initial state in
$\RRR^d$ to have support on the unit cube centered at the
corresponding lattice point. Convergence of the
$\PPP^{(n)}_{\rho,n}$ for the walk on $\RRR^d$ then implies the
convergence considered in \cite{K2,GJS} for walks on $\ZZZ^d$.

\begin{remark}\label{R2}
The scaling by $1/n$ relates the description of the asymptotic
distribution of the quantum walk to the equation of motion for a
quantum particle whose wavefunction propagates linearly 
in time, with a constant (but random) velocity vector
that has the distribution $\PPP_\rho$. Hence, 
a physical interpretation of
the commuting
bounded operators $V_1,\dots,V_d$ corresponding to the spectral
measure $P$ (cf.\ Theorem~\ref{theDualTheorem}) 
is that they are 
commuting observables giving the asymptotic velocity vector of
the quantum random walk. Note, however, that although 
this describes the asymptotic distribution, it does not describe the
quantum random walk precisely; indeed, the observables corresponding
to the positions at different times
do not commute
and hence one cannot regard the quantum random walk as a stochastic
process, since that would mean one could 
introduce a common
probability space for the positions at different times.
\end{remark}

\section{Proof of Theorem~\ref{theTheorem}}

Theorem~\ref{theTheorem} is formulated in the Schr\"odinger
picture, where the evolution is applied to the state of the
system. In the Heisenberg picture, where the the state of the
system remains constant and the physical observables evolve, the
quantum walk concerns the map
\[
     X \ \longmapsto \ X,\ U^* X U,\  U^{*2} X U^2,\ U^{*3} X U^3,\ \ldots
\]
from bounded operators $X$ on $\HH$ to sequences of bounded
operators.  We are interested in operators $X$ that represent
physical observables of position alone.  These are operators of
the form $m[f(x)]\otimes I_s$, where $m[f(x)]$ denotes the
operator $\psi(x) \longmapsto f(x)\psi(x)$ on $L^2(\RRR^d)$.
Theorem~\ref{theTheorem} is the consequence of a stronger, dual
formulation in the Heisenberg picture.  In the following, the
notation $\slim$ designates the limit in the strong operator
topology, the topology in which nets of operators converge if and
only if they converge pointwise as functions on $\HH$.

\begin{theorem}
\label{theDualTheorem} Let $U$ be a unitary operator on $\HH =
L^2(\RRR^d)\otimes \CCC^s$ of the form \eqref{step}.

There exists a spectral measure $P(dx)$ on $\RRR^d$ and a
corresponding family
of commuting bounded self-adjoint
operators $\dotsv$ such that
\begin{equation}
\label{heisenberg}
     \slim_{n \rightarrow \infty}
         U^{*n} ( m[f(x/n)]\otimes I_s) U^n
         \ = \ \int_{\RRR^d} f(x)P(dx)
\ = \ f(\dotsv).
\end{equation}
for all $f \in C_b(\RRR^d)$.
\end{theorem}

Theorem~\ref{theTheorem} follows from Theorem~\ref{theDualTheorem}
and the observation that if $\rho$ is a pure state, then
\begin{eqnarray}
       \lim_{n \rightarrow \infty}\PPP^{(n)}_{\rho,n}f(x)
       & = &
       \lim_{n \rightarrow \infty}
       \Tr\bigl( U^n \rho U^{*n} ( m[f(x/n)]\otimes I_s) \bigr)
       \nonumber 
\\
       & = &
       \Tr \Big(  \rho\ \bigl(\slim_{n \rightarrow \infty}
               U^{*n} ( m[f(x/n)]\otimes I_s) U^n  \bigr) \Big)
       \nonumber 
\\
       & = &
       \Tr \Big( \rho \int f(x)P(dx) \Big) \ = \ \int f(x) \Tr( \rho
       P(dx))
       \label{aa}
\end{eqnarray}
for any $f \in C_b(\RRR^n)$.  The
validity extends to general $\rho$ by a density argument.

Thus, it suffices to prove Theorem~\ref{theDualTheorem}.  Here is
the plan of the proof:  Note that, for each $n$, the map
\begin{equation}
\label{unitaryRepresentation}
    {\mathrm {\bf U}}_n:\omega \longmapsto {U^*}^n
 (m[e^{i x\cdot \omega/n}]\otimes I_s) U^n
\end{equation}
 is a strongly continuous unitary representation of $\RRR^d$ on
$L^2(\RRR^d,\CCC^s)$, i.e., ${\mathrm {\bf U}}_n$ is a group
homomorphism from $\RRR^d$ to the unitary operators on $\HH$ such
that ${\mathrm {\bf U}}_n(\omega)\psi$ is continuous in $\omega$
for each $\psi \in \HH$.   We will prove that the ${\mathrm {\bf
U}}_n(\omega)$ converge pointwise in the strong operator topology
as $n \longrightarrow \infty$ and we will identify the limit as a
uniformly continuous unitary representation of $\RRR^d$. By
Stone's spectral theorem there exists a spectral measure $P(dx)$
on $\RRR^d$ such that
\[
   \slim_{n\rightarrow\infty} U^{*n} (m[e^{i\omega\cdot x/n}]\otimes I_s) U^n
   \ = \ \int_{\RRR^d} e^{i\omega \cdot x} P(dx)
\]
for all $\omega \in \RRR^d$.  This proves that
\begin{equation}
   \slim_{n\rightarrow\infty} U^{*n} (m[f(x/n)]\otimes I_s) U^n
   \ = \ \int f(x) P(dx)
\label{reallyImportant}
\end{equation}
for all $f \in \spann\{ e^{i\omega\cdot x} : \omega \in \RRR^d\}$.
Finally, to complete the proof of Theorem~\ref{theDualTheorem}, we
will extend the convergence in (\ref{reallyImportant}) to all
functions $f \in C_b(\RRR^d)$.

We begin by showing that ${\Ub}_n(\omega)$ converges strongly at
each $\omega \in \RRR^d$.  To do this, we will first prove that
the infinitesimal generators of these unitary groups converge on a
dense subset of $\HH$ to a bounded skew-Hermitian operator, and
then invoke the Trotter--Kato Theorem.

 For arbitrary but fixed $\omega \in \RRR^d$, the
one-parameter unitary group $\{\Ub_n(t\omega)\}_{t\in \RRR}$ has
infinitesimal generator
\begin{equation}
\label{generators}
 \GG_n \ = \ \tfrac{1}{n}{U^*}^n (m[i\omega\cdot
x]\otimes I_s) U^n.
\end{equation}
Observe that
if $D(\omega)$ is the operator on $\CCC^s$
represented by the diagonal matrix
\[
   D(\omega) \ = \ \left[
\begin{array}{cccc}
   \omega \cdot v_1  & 0  & \cdots & 0 \\
 0 &   \omega \cdot v_2   & \cdots & 0 \\
 \vdots & \vdots & \ddots      & \vdots \\
 0 & 0 &  \cdots &  \omega\cdot v_s  \\
\end{array}
\right],
\]
then, for any $f\in L^2(\RRR^d)$,
\begin{equation*}
T^* (m[\omega\cdot x]\otimes I_s) T(f\otimes e_j) = m[\omega\cdot
(x+v_j)]\otimes I_s(f\otimes e_j) = \Bigl(m[\omega\cdot x]\otimes
I_s+I\otimes D(\omega)\Bigr)(f\otimes e_j)
\end{equation*}
and hence
\begin{eqnarray}
       U^* (m[i\omega\cdot x]\otimes I_s) U
& = &  (I\otimes C^*)  T^*(m[i\omega\cdot x]\otimes
       I_s)T (I\otimes C)  \nonumber \\
       & = &
       m[i\omega\cdot x]\otimes I_s
       \ + \ i(I \otimes C^*D(\omega) C)
       \label{forInduction} \ .
\end{eqnarray}
Applying \eqref{forInduction} recursively in \eqref{generators}
shows that
\begin{equation}
\label{generators2}
 \GG_n \ = \ \tfrac1n  m[i\omega\cdot x]\otimes I_s
 \ + \ i\ \frac1n \sum_{j=0}^{n-1} U^{*j}( I \otimes C^*D(\omega)
 C) U^j\ .
\end{equation}
The first term on the right-hand side of \eqref{generators2} is an
unbounded multiplication operator; as $n \longrightarrow \infty$
these operators converge to the zero operator on the dense subset
of $L^2(\RRR^d,\CCC^s)$ consisting of functions of bounded
support.  The second term on the right-hand side of
\eqref{generators2} is a bounded operator; we will prove that
these bounded operators converge strongly as $n \longrightarrow
\infty$.

We take Fourier transforms to identify the limit of the operators
\begin{equation}
\label{identifyMyLimit}
 \frac{1}{n}\sum_{j=0}^{n-1} U^{*j} ( I
\otimes C^*D(\omega) C) U^j\ .
\end{equation}
 Let $\FF$ denote the Fourier transform
\[
\FF(\psi)(k) \ = \   (2\pi)^{-d/2} \int_{\RRR^d} \psi(x) e^{-i k
\cdot x} dx
\]
on $\RRR^d$.  This is a unitary transformation from $L^2(\RRR^d)$
onto an isomorphic space $\widehat{L^2}(\RRR^d)$ with inverse
\[
\FF^*(\phi)(x) \ = \   (2\pi)^{-d/2} \int_{\RRR^d} \phi(k) e^{i k
\cdot x} dk \ .
\]
The spaces $L^2(\RRR^d)\otimes \CCC^s$ and
$\widehat{L^2}(\RRR^d)\otimes \CCC^s$ are mapped onto one another
via the unitary transformations $\FF\otimes I_s$ and $\FF^*\otimes
I_s$.  We will denote $\FF\otimes I_s$ by $ \FFI $ from now on.
The operator $U$ of (\ref{step}) is unitarily equivalent to the
operator $\widehat{U} = \FFI U \FFI^*$ on $\widehat{\HH}$.  We
will think of $\widehat{\HH}$ 
(the momentum space of the system)
as $L^2(\RRR^d,\CCC^s)$, 
that is, we represent vectors in
$\widehat{\HH}$ by square-integrable column-vector valued
functions on $\RRR^d$.  Then the operator $\widehat{U}$ may then
be represented by a ``matrix-multiplication operator''
\[
        (\widehat{U}\phi)(k) \ = \ \widehat{U}_k  \phi(k)
\]
with
\begin{equation}
\label{Uhat-k}
 \widehat{U}_k
\ = \  \left[
\begin{array}{cccc}
 e^{-ik \cdot v_1} & 0 & \cdots  & 0 \\
 0 & e^{-ik \cdot v_2} & \cdots  & 0 \\
 \vdots & \vdots &  \ddots & \vdots \\
 0 & 0 & \cdots &  e^{-ik \cdot v_s} \\
\end{array}
\right]
 \left[
\begin{array}{ccccc}
 C_{11} & C_{12} & \cdots &  C_{1s} \\
 C_{21} & C_{22} & \cdots &  C_{2s} \\
 \vdots & \vdots & \ddots & \vdots \\
 C_{s1} & C_{s2} & \cdots  & C_{ss} \\
\end{array}
\right]
\end{equation}
where the $C_{jk}$ denote the matrix elements of the coin toss
operator $C$.  The Fourier transform of the operator
$\frac{1}{n}\sum_{j=0}^{n-1} U^{*j} ( I \otimes (C^*\Domega C) )
U^j$ is the matrix-multiplication operator
\begin{equation}
\label{matrixMultiplicationOperator}
      \phi(k) \ \longmapsto \ \frac1n \sum_{j=0}^{n-1}
         \widehat{U}_k^{*j}  C^*D(\omega) C  {\widehat{U}_k}^j \ \phi(k) \ .
\end{equation}
For fixed $k \in \RRR^d$ the map $X \longmapsto \widehat{U}_k^* X
\widehat{U}_k$ is a unitary operator on the space of $s \times s$
matrices endowed with the inner product $\langle Y, X \rangle =
\tr(Y^*X)$. The Mean Ergodic Theorem implies that the matrices in
\eqref{matrixMultiplicationOperator} converge as $n
\longrightarrow \infty$ to the orthogonal projection of
$C^*\Domega C $ onto the subspace of $s \times s$ matrices that
commute with $\widehat{U}_k$.  That is,
\begin{equation}
\label{MET}
      W_k(\omega) \ = \ \lim_{n \rightarrow \infty} \frac1n \sum_{j=0}^{n-1}
         \widehat{U}_k^{*j} ( C^*D({\omega}) C ) {\widehat{U}_k}^j
\end{equation}
is the unique matrix that commutes with $\widehat{U}_k$ and
satisfies
\begin{equation}
\label{Vk}
        \tr( N^* W_k(\omega) ) \ = \ \tr( N^* C^*D(\omega) C )
\end{equation}
for all matrices $N$ that commute with $\widehat{U}_k$.   Since
$C^*D(\omega) C$ is Hermitian and since $N$ commutes with
$\widehat{U}_k$ if and only if $N^*$ commutes with
$\widehat{U}_k$, taking adjoints in \eqref{Vk} shows that the
matrices $W_k(\omega)$ are Hermitian. From \eqref{Uhat-k} and
\eqref{MET} it is seen that $W_k(\omega)$ is a limit of continuous
functions of $k$, hence it is a measurable function of $k$.
Moreover, the operator norm of each matrix on the right-hand side
of \eqref{MET} does not exceed the operator norm of $D(\omega)$,
which is independent of $k$, and the limit $W_k(\omega)$ enjoys
the same uniform bound.  Thanks to these bounds, Lebesgue's
Dominated Convergence Theorem implies that the operators
\eqref{matrixMultiplicationOperator} converge strongly to the
matrix-multiplication operator $\WW(\omega)$ defined by
\begin{equation}
\label{VV}
       (\WW(\omega)\phi)(k) \ = \ W_k(\omega) \phi(k)
\end{equation}
on $\widehat{\HH}$, a bounded Hermitian operator.  Transforming
back to operators on $\HH$, we conclude that the operators
\eqref{identifyMyLimit} converge strongly to
\begin{equation}\label{VV2}
 \VV(\omega) =\FFI^* \WW(\omega)\FFI.
\end{equation}

We have now shown that the generators \eqref{generators2} converge
to $i \VV(\omega)$
on a dense subset of $\HH$.  By the
Trotter--Kato Theorem  \cite[Theorem 4.5]{Pazy}, the operators
${\Ub}_n(t\omega)$ converge strongly to $e^{it\VV(\omega)}$ for
each $t \in \RRR$.  In particular, ${\Ub}_n(\omega)$ converges
strongly to $e^{i\VV(\omega)}$, i.e.,
\[
   \slim_{n\rightarrow\infty} U^{*n} (m[e^{i\omega\cdot x/n}]\otimes I_s) U^n
   \ = \ e^{i\VV(\omega)}\ .
\]

It follows from \eqref{MET}, \eqref{VV} and \eqref{VV2} that
$\VV(\omega)$ depends linearly on $\omega\in\RRR^d$; in
particular, $\omega\longmapsto\VV(\omega)$ is continuous.
Moreover, $\omega \longmapsto e^{i\VV(\omega)}$ is a $d$-parameter
group of unitary operators on $\HH$ because each $\omega
\longmapsto \Ub_n(\omega)$ is. It follows that $e^{i\VV(\omega)}$
is a uniformly continuous $d$-parameter unitary group.  Stone's
spectral theorem for one-parameter unitary groups can be
generalized to $d$-parameter unitary groups: for any strongly
continuous $d$-parameter unitary group ${\Ub}(\omega)$ there
exists a spectral measure $P(dx)$ on $\RRR^d$ such that
\[
      {\Ub}(\omega) \ = \ \int_{\RRR^d} e^{i
      \omega\cdot x} P(dx)
\]
(see, e.g., \cite[Chapter 2.4]{Helson}). For one-parameter groups
$e^{i\omega V}$, $P(dx)$ is just the spectral measure associated
to the self-adjoint operator $V$,
and in general $P$ is the
spectral measure of a $d$-tuple $(\dotsv)$ of commuting bounded
operators.  Letting $P(dx)$ denote the spectral measure associated
to $e^{i\VV(\omega)}$, what we have shown up to this point is that
\begin{equation}
   \slim_{n\rightarrow\infty} U^{*n} (m[e^{i \omega \cdot x/n}]\otimes I_s) U^n
   \ = \ e^{i\VV(\omega)} \ = \ \int_{\RRR^d} e^{i \omega \cdot x} P(dx)
\label{exponentials}
\end{equation}
for all $\omega \in \RRR^d$.

To complete the proof of Theorem~\ref{theDualTheorem}, we have to
extend to all functions in $C_b(\RRR^d)$ the convergence
\eqref{exponentials} just established for functions
$e^{i\omega\cdot x}$. By linearity, the assertion of the theorem
holds for all  $f \in \spann\{ e^{i\omega\cdot x}: \omega \in
\RRR^d\}$.  Let $f$ be an arbitrary but fixed function in
$C_b(\RRR^d)$ and define
\[
  \Lx[f,n](\psi) \ = \ U^{*n}(m[f(x/n)] \otimes I_s) U^n \psi
\]
on $\HH$.  We want to prove that
\[
  \lim_{n \rightarrow \infty} \Lx[f,n](\psi) \ = \ \int f(x)
  P(dx)\psi
\]
for all $\psi\in \HH$. The sequence $\{\Lx[f,n]\}$ is
equicontinuous on $\HH$ thanks to the uniform bound
\[
   \| U^{*n}(m[f(x/n)] \otimes I_s) U^n \| \ = \  \| f \|_{\infty}\ ,
\]
so it suffices to prove the convergence of the sequence
$\{\Lx[f,n]\}$ on a dense subset of $\HH$.  We will show that
$\Lx[f,n](\psi)$ converges for all $\psi$ that have bounded
support.

We continue to represent members of $\HH$ by functions in
$L^2(\RRR^d,\CCC^s)$.  Let $\psi\in \HH$ have bounded support, in
the sense that there exists $r>0$ such that $\psi(x)$ equals ${\bf
0}\in \CCC^s$ when $|x|>r$.  We claim that $\Lx[f,n]\psi$ tends to
a limit.  We may assume that $\psi$ has norm $1$ without loss of
generality. Let
\[
    v \ = \ \max_{1\le j \le s} \big\{|v_j|\big\} \ ,
\]
where the $v_j$ are the translation vectors appearing in the
definition \eqref{conditionalTranslation} of $T$.    Then
$U^n\psi$ is supported on the ball of radius $r+nv$ in $\RRR^d$.
For any $g \in C_b(\RRR^d)$,
\begin{eqnarray}
  \big\| \Lx[f,n]\psi  - \Lx[g,n]\psi \big\|
  & = &
  \big\| U^{*n} ((m[f(x/n)]-m[g(x/n)]) \otimes I_s) U^n \psi  \big\| \nonumber \\
  & \le &
  \sup_{|x|< r+v}\{ |f(x)-g(x)|\}
\label{uniform}
\end{eqnarray}
for all $n$ because of the way that $f$ and $g$ are scaled.  By
the Stone--Weierstrass Theorem, the linear span of $\{
e^{i\omega\cdot x} : \omega \in \RRR^d\}$ is dense in the space of
continuous functions on any compact subset of $\RRR^d$, so $f(x)$
can be uniformly approximated within arbitrary $\epsilon > 0$ on
the ball of radius $r+v$ by some function $g_{\epsilon} \in \spann
\{ e^{i\omega\cdot x} : \omega \in \RRR^d\}$.  We know that each
sequence $ \Lx[g_{\epsilon},n]\psi $ converges as $n
\longrightarrow \infty$.  On the other hand, $
\Lx[g_{\epsilon},n]\psi $ is within $\epsilon$ of $\Lx[f,n]\psi$
uniformly in $n$ by \eqref{uniform}.  It follows that
$\{\Lx[f,n]\psi\}$ converges as $n \longrightarrow \infty$. The
limit is
a continuous function of $f$ and hence must agree
everywhere with $\int f(x) P(dx)\psi=f(\dotsv)$.  This proves
\eqref{heisenberg} for arbitrary $f \in C_b(\RRR^d)$ and concludes
the proof of Theorem~\ref{theDualTheorem}.

\section{The limit distribution}

In special cases there is a nice characterization of the limit
distribution $\PPP_{\rho}$ of Theorem \ref{theTheorem}
in terms of the
distribution of a certain random element of $\Omega=\RRR^d \times
\{1,2,\ldots, s \}$.

 Let $\whrho=\FFI\rho\FFI^*$ be the density
operator on $\widehat{\HH}$ corresponding to $\rho$. This can be
regarded as an integral operator $ \whrho f = \int_{\RRR^d}
\whrho(k,k')f(k')\,dk'$ with an $s\times s$ matrix valued kernel $
\whrho(k,k')$. Since $\whrho$ is trace class, the diagonal $
\whrho(k,k)$ is well-defined a.e.\ and integrable.
  By Theorem \ref{theTheorem} and \eqref{aa},
  \eqref{exponentials}, we have
\eqref{VV2},
\begin{equation*}
\PPP_{\rho}[e^{i\omega \cdot x}]
 =
\Tr\Bigl(\rho \int_{\RRR^d}  e^{ i \omega \cdot x} \, dP(x) \Bigr)
= \Tr\Bigl(\rho e^{i\VV(\omega)} \Bigr) = \Tr\Bigl(\whrho
e^{i\WW(\omega)} \Bigr) =
    \int_{\RRR^d}
  \tr \big[ e^{iW_k(\omega)} \whrho(k,k) \big] \,d k.
\end{equation*}
 This shows that the characteristic function of the limit
$\PPP_{\rho}$ in Theorem \ref{theTheorem} is
\begin{equation}\label{hamletAlex}
   \PPP_{\rho}[e^{i\omega \cdot x}] \ = \ \int_{\RRR^d}
  \tr \big[ e^{iW_k(\omega)} \whrho(k,k) \big] \,d k,
\end{equation}
with $W_k$ given by \eqref{MET} and \eqref{Vk}.

We can better identify the limit measure $\PPP_{\rho}$ in cases
where the matrices $\hU_{k}$ have $s$ distinct eigenvalues
$\lambda_1(k),\ldots,\lambda_s(k)$ for a.e.\ $k\in\RRR^d$. At a
point $k$ where there are $s$ distinct eigenvalues, let
$\psi_j(k)$ be a normalized eigenvector of $\hU_k$ with eigenvalue
$\lambda_j(k)$, and let $\pkj=\psi_j(k)\psi_j(k)^*$ be the
corresponding orthogonal projection onto the eigenspace. We may
assume that the eigenvalues are numbered such that $\lambda_j$ are
continuous and differentiable in a neighborhood of $k$ for each
$j$ \cite{Kato}[II-5.4], and we then define
\begin{equation}
  \label{piAlex}
 \pi(k,j) \ = \
            i\overline{\lambda_j(k)}\,\nabla \lambda_j(k),
\end{equation}
where $\nabla$ is the gradient with respect to $k\in\RRR^d$. Since
$|\lambda_j(k)| \equiv 1$, the functions
$i\overline{\lambda_j(k)}\nabla\lambda_j(k)$ are real-valued, and
$\pi$ is an a.e.\ defined map from $\RRR^d \times \{1,2,\ldots, s
\}$ to $\RRR^d$.

\begin{lemma}
Suppose that $\hU_{k}$ has $s$ distinct eigenvalues at almost
every $k$. Then,
\begin{equation}
  \label{wkjAlex}
  W_{k}(\omega) \ = \ \sum_{j=1}^s \bigl(\omega\cdot\pi(k,j)\bigr) \pkj
\end{equation}
and thus
\begin{equation}
\label{characteristicBecomesAlex} \PPP_{\rho}[e^{i\omega \cdot x}]
\ = \ \int_{\RRR^d} \sum_{j=1}^s e^{i \omega \cdot \pi(k,j) } \tr[
\whrho(k,k) \pkj]   dk \ .
\end{equation}
\end{lemma}

\begin{proof}
  We may assume that the eigenvectors $\psi_j$ are differentiable in a
  neighborhood of $k$ \cite{Kato}[II-5.4].
Taking the directional derivative $\ddo=\omega\cdot\nabla$ of
  $\hU_k\psi_j(k)=\lambda_j(k)\psi_j(k)$ we obtain, since
\eqref{Uhat-k} yields $\ddo\hU_k = -iD(\omega)\hU_k$,
\begin{equation}
  \label{duAlex}
-iD(\omega)\hU_k\psi_j(k) +\hU_k\ddo\psi_j(k)
=\bigl(\ddo\lambda_j(k)\bigr)\psi_j(k) +\lambda_j(k)\ddo\psi_j(k).
\end{equation}
Further,
\begin{equation*}
\psi_j(k)^*\hU_k =\bigl(\hU_k^*\psi_j(k)\bigr)^*
=\bigl(\overline{\lambda_j(k)}\psi_j(k)\bigr)^*
=\lambda_j(k)\psi_j(k)^*
\end{equation*}
and thus \eqref{duAlex} implies
\begin{equation}\label{du1Alex}
\psi_j(k)^*\bigl(-iD(\omega)\hU_k\bigr)\psi_j(k)
=\psi_j(k)^*\bigl(\ddo\lambda_j(k)\bigr)\psi_j(k)
=\ddo\lambda_j(k).
\end{equation}
Since $W_k(\omega)$ commutes with $\hU_k$ it is of the form
$\sum_j a_j\pkj$.
Taking $N=N^*=\psi_j(k)\psi_j(k)^*$ in \eqref{Vk} yields, using
$C^*D(\omega)C=\hU_k^*D(\omega)\hU_k$, \eqref{du1Alex} and
\eqref{piAlex},
\begin{align*}
  a_j
& =\Tr\Bigl(\psi_j(k)\psi_j(k)^* W_k(\omega)\Bigr)
=\Tr\Bigl(\psi_j(k)\psi_j(k)^* C^*D(\omega)C\Bigr) =\psi_j(k)^*
\hU_k^*D(\omega)\hU_k\psi_j(k)
\\
&=\overline{\lambda_j(k)}\psi_j(k)^* D(\omega)\hU_k\psi_j(k)
=i\overline{\lambda_j(k)}\,\ddo \lambda_j(k) =\omega\cdot\pi(k,j).
\end{align*}
This gives \eqref{wkjAlex}, and \eqref{characteristicBecomesAlex}
then follows from \eqref{hamletAlex}.
\end{proof}

Let $\wt{\PPP}_{\rho}$ denote the probability measure
\begin{equation}\label{macbethAlex}
  \wt{\PPP}_{\rho}(dk, j)
\ = \ \tr[ \whrho(k,k) P_{k j}]\ dk
 \end{equation}
on $\Omega=\RRR^d \times \{1,2,\ldots, s \}$. Then the right-hand
side of (\ref{characteristicBecomesAlex}) is the characteristic
function of the induced probability measure $
\wt{\PPP}_{\rho}\circ\pi^{-1}$ on $\RRR^d$, which identifies this
probability measure as $\PPP_{\rho}$. In other words, if $Y$ is a
random element of $\Omega$ with the distribution
$\wt{\PPP}_{\rho}$, then $\PPP_{\rho}$ is the distribution of the
random variable $\pi(Y)$. We conclude with an interesting special
case:
\begin{corollary}\label{Cor}
Suppose that the initial state is is of the form
  $\rho=\rho_0\otimes \frac1s I_s$, a tensor product of a position
density operator $\rho_0$ on $L^2(\RRR^d)$ with a maximally mixed
coin state. Suppose further that the matrices $\hU_{k}$ have $s$
distinct eigenvalues $\lambda_1(k),\ldots,\lambda_s(k)$ almost
everywhere on  $\RRR^d$. Then the limit $\PPP_{\rho}$ is the
distribution of the random variable $\pi(Y_0,Z)$, where $\pi$ is
given by \eqref{piAlex}, $Y_0$ is a random vector in $\RRR^d$ with
density function $\Tr \widehat{\rho_0}(k,k)$ and $Z$ is uniformly
distributed on $\{1,\dots,s\}$, with $Y_0$ and $Z$ independent.
\end{corollary}

\begin{proof}
We have $\whrho=\widehat{\rho_0}\otimes \frac1s I_s$ and, for the
kernel, $\whrho(k,k)=\widehat{\rho_0}(k,k) \frac1s I_s$. Since
each $P_{k j}$ has rank 1, \eqref{macbethAlex} shows that
\begin{equation*}
\wt{\PPP}_{\rho}(dk, j) \ = \ \tfrac1s \widehat{\rho_0}(k,k) dk.
\end{equation*}
This is a product measure and thus, if  $Y$ above is written as
$(Y_0,Z)$, then $Y_0$ and $Z$ are independent with the stated
distributions.
\end{proof}

Note that in Corollary \ref{Cor}, the definition of the quantum
walk, as given by \eqref{conditionalTranslation} and \eqref{step},
affects only $\pi$, while the initial state affects only the
distribution $Y_0$.

\begin{remark}
As explained at the end of Section \ref{S:gen}, the results
transfer to quantum random walks on $\ZZZ^d$ too. It can be
verified, using the method described there, that in this case the
limit distribution is described by the formulas above, but with
$k\in \KKK^d$, the dual group (a $d$-dimensional torus). This
generalises results in \cite{GJS}.  In particular, if the initial
position is $0\in\ZZZ^d$ with a maximally mixed coin state,
Corollary~\ref{Cor} holds with $Y_0$ uniformly distributed on
$\KKK^d$.
\end{remark}

\bigskip

\noindent {\bf Acknowledgments.}  A.G. is supported by the
Austrian START project ``Nonlinear Schr\"odinger and quantum
Boltzmann equations'' of Norbert J. Mauser (contract Y-137-Tec).
S.J. was supported by the Swedish Royal Academy of Sciences, the
London Mathematical Society and Churchill College, Cambridge.
P.F.S. acknowledges the EU (grant HPRN-CT-2002-002777) and Prof.\
Joseph Avron for support. P.F.S. thanks Geoffrey Grimmett, Netanel
Lindner and Terry Rudolph for helpful comments and discussions.

\end{document}